\documentclass[11pt,amsmath,amssymb,nofootinbib]{revtex4}
\usepackage{amsmath}
\usepackage{amsfonts}
\usepackage{amssymb}
\usepackage[dvips]{graphicx}

\begin{document}

\newcommand\be{\begin{equation}}
\newcommand\ee{\end{equation}}
\newcommand\ba{\begin{eqnarray}}
\newcommand\ea{\end{eqnarray}}
\newcommand\bseq{\begin{subequations}} 
\newcommand\eseq{\end{subequations}}
\newcommand\bcas{\begin{cases}}
\newcommand\ecas{\end{cases}}
\newcommand{\p}{\partial}
\newcommand{\f}{\frac}
\newcommand{\nn}{\nonumber \\}
\def\tr{{\rm Tr}}

\title{Towards inhomogeneous loop quantum cosmology:\\ triangulating Bianchi IX with perturbations}

\author{Antonino Marcian\`o}
\address{Centre de Physique Th\'eorique,\\
Case 907 Luminy, 13288 Marseille, EU\\
E-mail: antonino.marciano@cpt.univ-mrs.fr}

\begin{abstract}
This brief article sums up results obtained in arXiv:0911.2653, which develops a constrained $SU(2)$ lattice gauge theory in the ``dipole'' approximation. This is a further step toward the issue of a (inhomogeneous) loop quantum cosmology and its merging into loop quantum gravity.

\end{abstract}

\maketitle
\bigskip

Loop quantum cosmology (LQC) is an application of quantisation techniques adopted within loop quantum gravity (LQG) \cite{Rovelli_book, Thiemann_book} to particular contexts in which external symmetries reduce this latter theory to a finite (quantum dynamical) system of variables. In these cases singularities have been successfully removed and implications to dynamics have been studied at different levels \cite{Ash_Big_Bounce}. Although the great successes gained by investigations within LQC, many points need still to be addressed in a more detailed analysis. For instance, main dynamical effects have been introduced into {\it effective classical equations}, in order to avoid ``interpretational'' problems originated by the quantum theory. Despite these efforts, it is not clear how the link between LQC models and the full theory is obtained in the dynamics.

This letter relies on the collaboration carried out in Ref.~\cite{BMR} (which I refer to for notations). It develops a constrained $SU(2)$ lattice-gauge theory applied to cosmology\cite{Rovelli&Vidotto1} making use of a Born-Hoppenheimer approximation method which treats homogeneous cosmological degrees of freedom (d.o.f.) as ``heavy ones'', and inhomogeneous d.o.f as ``light ones''. The main results of Ref.~\cite{BMR} are, at the classical level, the derivation of effective $k=1$ FRW and Bianchi IX equations directly from the constrained $SU(2)$ lattice gauge theory considered, and the occurrence of a perturbative way of dealing with inhomogeneities in the Bianchi IX background. 

For a generic  cellular complex triangualtion $\Delta_n$ on $S^3$ one can  associate a group element $U_f\in SU(2)$ and a $su(2)$ algebra element $E_f=E_f^I\tau_I$ to each oriented triangle $f$. (I remind that for a triangulation making use of $n$ tetrahedra $t$ there are $2n$ triangles $f$.) 
The phase space is then the cotangent bundle of $SU(2)^{2n}$ with its natural symplectic structure:
$\{  U_f,U_{f'} \} = 0,  \,\,
\{  E_f^I,U_{f'} \} =  \delta_{\!f\!f'} \ \tau^I U_{f}, \,\, {\rm and}  \,\,
\{  E_f^I,E_{f'}^J \} = - \  \delta_{\!f\!f'}\  \epsilon^{IJK} E_f^K.   $
The dynamics is encoded in the two sets of constraints (here we consider Euclidean gravity and set the Immirzi parameter $\beta=1$) namely
\begin{equation} \nonumber
G_t \equiv  \sum_{f\in t} E_f \sim 0, \quad  
C_t \equiv  V^{-1}_t \sum_{ff'\in t} {\rm tr}[U_{\!f\!f'}E_{f'}E_{f},] \sim 0\,, \quad {\rm with} \,\,\,\,\,V_t^2=  {\rm tr}[E_fE_{f'}E_{f'\!'}]\,.
\end{equation}

In order to analyse the features of this SU(2) lattice gauge-theory endowed with $S^3$ topology, we adapted the analysis in Ref.~\cite{Rovelli&Vidotto1} to the case of the Bianchi IX model, which is the most general homogenous space with the topology of $S^3$. In terms of the homogeneous reference triad fields, the 3D spatial slices are characterised by the Maurer-Cartan (flat) connections $\omega=g^{-1}dg=\omega^{I} \tau_I$ (fulfilling $d\omega^{I} = \frac{1}{2} \epsilon^{\!I} _{\,\,\,JK} \, \omega^{J} \wedge \omega^{K}$) and by the left-invariant vector fields $e^I$ taking values in $su(2)$, which are dual to $\omega^J$ via $e^a_I\omega^J_a=\delta^J_I$ and fulfill the Lie brackets $[e_{I},\,e_{J} ]= -\epsilon_{\,\,\,IJ}^{K} \,e_{K}$.
By means of the first order formalism, the Ashtekar-Barbero variables are expressed as $A^I_a=c_{I}\, \omega^{I}_a$ and as $E^a_I=p^{I}\,\omega\, e^a_{I}$, in which $\omega$ stands for the determinant of $\omega^I_a$. 

In the dipole (two tetrahedra) approximation of $S^3$, the model is defined by the dual graph $\Delta_2^*=\!\!\!\!\begin{picture}(20,5)
\put(20,4) {\circle{15}}
\put(12,4) {\circle*{2}} 
\put(28,4) {\circle*{2}}  
\qbezier(12,4)(19,13)(28,4)
\qbezier(12,4)(19,-5)(28,4)
\end{picture} \,\,\,\,\,\,$, which in turns specifies the cellular complex triangulation of the model.
The $SU(2)$ symmetry structure enters twice in our description: in adding inhomogeneities and in discretising Ashtekar-Barbero variables. Indeed 
the flux of the electric field rewrites exactly as $E_f=p^I\,\omega^I_f\tau_I$, while the holonomy around the the dual links of $\Delta_2^*$ can be approximated (for small curvature, and hence for small $c^I$) as $U_f=\exp\,c^I\,\omega^I_f\tau_I$ (see Ref.~\cite{BMR}). The $\omega^I_f$, which appear in the two formulas above, are circuitations along the dual links of $\Delta_2^*$ and represent the flux of the Plebanski two-form across triangles of $\Delta_2^*$:
\begin{equation}\nonumber
\omega_{f}^I \equiv  \int_f \Sigma^I = \frac{1}{2}\,\int_f  \epsilon^I\,_{JK}\,\omega^J \wedge \omega^K = 
\int_f  d \omega^I = \oint_{\partial f}  \omega^I. 
\end{equation}
As a consequence 
 the equality $\sum_{f\in t} \omega_{f}^I =0$, which ensures the Gau\ss\, constraint for the $E_f$ variables, holds exactly for each tetrahedron $t$.

In the homogeneous isotropic case, we obtained a ``natural'' effective Hamiltonian constraint by using the discretized scalar constrained of the theory without need of a ``polymerization'' of the classical model: \begin{equation} \nonumber
\tilde C=\frac{17}{6}p^2\left[\cos(c-\alpha)-1\right]\approx0\,,
\end{equation}
where $\alpha$ takes into account the contribution of the $S^3$ curvature to the holonomy in the {\it ansatz} $U_f=\exp\,(c+\alpha)\,\omega^I_f\tau_I$. The value of $\alpha$, such that $\cos\alpha=(9-\sqrt{17})/8$, is fixed by requiring matching with ordinary classical dynamics for 
 $|c|\ll1$.

The quantization of the homogeneous and isotropic sector of the model is straightforward.
The kinematic Hilbert space $\mathcal{H}_{iso}$ of the theory is $L^2(S^1,dc/4\pi)$ of square integrable functions on a circle, since the variable $c$ entering the definition of the holonomy multiplies a generator of a $U(1)$ subgroup of the compact gauge
group $SU(2)$. Eigenstates of $\hat{p}$ are labelled by integer $\mu$ and read $\langle c|\mu\rangle=e^{i\mu c/2}$, and wave functions $\psi(c)$ are decomposed in a Fourier series of eigenstates of $p$. 
The fundamental operators are 
$p$ and $\exp(i c/2)$ and their action reads $
p\, |\mu\rangle = \mu/2\ |\mu\rangle$ and $\exp(i c/2)\,|\mu\rangle=|\mu+1\rangle$. 
Thus the quantum constraint operator rewrites as a difference equation for the coefficients $\psi_\mu=\langle c|\,\mu\rangle$ (see Ref.~\cite{BMR} for details on the classical and quantum description of the Bianchi IX model).

Inhomogeneous perturbations to the (classical) Bianchi IX Universe can be described in terms of the $SU(2)$ symmetry structure \cite{Regge&Hu}. Indeed tensorial perturbations to the three metric $q^0_{ab}(x,t)$ can be recast in terms of a matrix of scalar perturbations to the triadic projection of the metric. If $h_{ab}(x,t)=q_{ab}(x,t)-q^0_{ab}(x,t)$ represents a general perturbation to $q^0_{ab}(x,t)$, one can expand its triadic projection $h_{IJ}(x,t)$ in terms of Wigner $D$-functions $D^j_{\alpha'\alpha}(g(x))$, where $g(x)$ stands for a group element of $SU(2)$ coordinatized by the spatial $x$ variables. Indeed, as the spatial slices of Bianchi IX have the $S^3$ topology, they are group homeomorphic to $SU(2)$. Thus a Peter-Weyl decomposition of $h_{IJ}(x,t)$ can be applied there where the sum over $j,\alpha$ labels contracted with the Wigner $D$ functions is now not performed (since the Einstein equations allow to decouple $j,\alpha$ modes from $\alpha'$ modes):
\begin{equation} \nonumber
h^{j\alpha}_{IJ}(x,t)= \sum_{\alpha' =-j}^{j} \, h_{IJ}^{j\alpha\alpha'}(t) \, D^j_{\alpha' \alpha}(g(x))\,,
\end{equation}
(see Appendix C of Ref.~\cite{BMR}). In the first order formalism, this corresponds to finding perturbations to $e^a_I(x)$ as $E^a_I(x,t)\!=\!e^a_I(x)+\psi^a_I(x,t)$, to projecting on co-triads $\omega_a^J$ and to expanding  in Wigner $D$-functions the $j,\alpha$ modes $\psi_{IJ}^{j\alpha}(x,t)$ within $\psi_{IJ}(x,t)=\sum_{j\alpha}\psi_{IJ}^{j\alpha}(x,t)$. In a similar way, co-triads $\tilde{\omega}^{I}_a(x,t)\!=\!\omega^{I}_a(x)+\varphi^{I}_a(x,t)$ expand by $\varphi_{ j\,\alpha}^{IJ}(x,t) =\,\sum_{ \alpha'=-j}^j \, \varphi^{IJ}_{j\alpha\alpha'}(t) \,  D^j_{\alpha' \alpha}\left(g(x)\right)$. Time-dependent expansion coefficients $\varphi^{IJ}_{j\alpha\alpha'}(t)$ and $\psi^{IJ}_{j\alpha\alpha'}(t)$ can be then used to capture inhomogeneous d.o.f. \cite{BMR}.

In the dipole model inhomogeneities are described by nine variables $\varphi^{I}_{\alpha}(t)$ plus nine $\psi^{I}_{\alpha}(t)$.  The Gau\ss\, constraint reduces d.o.f. from nine to six (see Ref.~\cite{BMR}): the number of d.o.f. of the dipole model is then recovered.
Discretization and quantization then follow the same lines sketched above: for co-triads $\tilde\omega_f^I=\omega^I_f+\sum_{j\alpha\alpha'}\,\varphi^{KL}_{j \alpha\alpha'}(t)\,  \phi_{f\,KL}^{I\,j\alpha\alpha'}$
and for triads $\tilde E_f^I=E^I_f+  2\sum_{j\alpha\alpha'}\, \psi_{IJ}^{j\alpha} \phi_{f, \alpha \alpha'}^J$, where $\phi_{f\,KL}^{I\,j\alpha\alpha'}=\int_f\epsilon^{I}_{\,JK}\, D^j_{\alpha\, \alpha'}\, \omega^J\wedge \omega^{L}$.

Perturbation to the Bianchi IX Universe in the dipole approximation introduces six inhomogeneous d.o.f. \cite{BMR}. Such a result must be yet generalized to the case of non vanishing Barbero-Immirzi parameter and to Lorentzian signature. This will be a first step in order to check at the quantum level the BKL conjecture, which says that classically, toward the singularity, the Universe is locally made of (lorentzian) Bianchi IX patches. 

This proposal represents a way for merging LQC in LQG. Progresses of this line of research, besides the points reminded above, aim at developing a spin foam evolution of this model, the main result of which would be to derive predictions on thermodynamics. For instance, in order to see accordance of this model with experimental data, we may look at anisotropies in the power spectrum of CMBR as induced by space-time quantum fluctuations.


\end{document}